\documentclass[preprint,authoryear]{elsarticle}

\usepackage{amssymb}
\usepackage{graphicx}
\usepackage{subfig}
\usepackage{aas_macros}
\usepackage{multirow}

\journal{Nuclear Instruments and Methods}

\begin{document}
\begin{frontmatter}

\title{X-ray Polarimetry: a new window on the high energy sky}

\author[INFN]{R.~Bellazzini}
\author[IASF]{F.~Muleri\corref{cor}}
\cortext[cor]{Corresponding author.}
\ead{fabio.muleri@iasf-roma.inaf.it}
\address[INFN]{INFN sez. Pisa, Largo B. Pontecorvo 3, I-56127 Pisa, Italy}
\address[IASF]{IASF/INAF, Via del Fosso del Cavaliere 100, I-00133 Roma, Italy}

\begin{abstract}
Polarimetry is widely considered a powerful observational technique in X-ray astronomy, useful to
enhance our understanding of the emission mechanisms, geometry and magnetic field arrangement of
many compact objects. However, the lack of suitable sensitive instrumentation in the X-ray energy
band has been the limiting factor for its development in the last three decades. Up to now,
polarization measurements have been made exclusively with Bragg diffraction at 45$^\circ$ or
Compton scattering at 90$^\circ$ and the only unambiguous detection of X-ray polarization has
been obtained for one of the brightest object in the X-ray sky, the Crab Nebula. Only recently, with
the development of a new class of high sensitivity imaging detectors, the possibility to exploit the
photoemission process to measure the photon polarization has become a reality. We will report on the
performance of an imaging X-ray polarimeter based on photoelectric effect. The device derives the
polarization information from the track of the photoelectrons imaged by a finely subdivided Gas
Pixel Detector. It has a great sensitivity even with telescopes of modest area and can perform
simultaneously good imaging, moderate spectroscopy and high rate timing. Being truly 2D it is
non-dispersive and does not require any rotation. This device is included in the scientific payload
of many proposals of satellite mission which have the potential to unveil polarimetry also in X-rays
in a few years.
\end{abstract}

\begin{keyword}
X-rays \sep Gas Detectors \sep Polarimetry

\PACS 29.40.Cs \sep 07.85.Fv \sep 95.55.Ka \sep 95.75.Hi

\end{keyword}

\end{frontmatter}

\section{Introduction}

X-ray polarimetry was born together with X-ray astronomy. First pioneering experiments were carried
out in the seventies with polarimeters based on Bragg diffraction at 45$^\circ$ or Compton
scattering at 90$^\circ$ on-board sounding rockets and first results were quite encouraging: already
Novick et al. \citep{Novick1972} reported a marginal yet significant detection of polarization in
the emission of the Crab Nebula, confirmed a few years later with high significance by the Bragg
polarimeter on-board OSO-8 \citep{Weisskopf1978}. This observation was favoured by the intense flux
of the source and the high degree of polarization, signature of synchrotron emission
($\mathcal{P}\approx$20\%), and, as a matter of fact, it has remained unique. Only upper limits were
derived for other astrophysical objects because of the combined effect of a lower flux and an
inferior polarization degree \citep{Long1979,Hughes1984}.

Unfortunately no other tool dedicated to X-ray polarimetry has been launched after OSO-8. The
Stellar X-ray Polarimeter on-board the Spectrum-X-Gamma mission, although the flight model was
ready and calibrated, was never put in orbit because of the collapse of Soviet System. The proposals
to include polarimeters on-board observatories like XMM or AXAF, which were the only opportunities
to have a sufficient collecting area, has never been carried out: instruments exploiting Bragg
diffraction or Compton scattering didn't look attractive because, while imaging and
spectroscopic devices promised an enormous increase of sensitivity, polarimetry would be limited to
a few bright sources even in the focus of these large telescopes. Moreover ``classical''
polarimeters were cumbersome because they need to be rotated around the direction of incident
photons.

Conversely, the lack of experimental feedback has not prevented the development of a rich literature
on the basis of which we expect that almost all sources in the X-ray sky should emit partially
polarized radiation \citep[for reviews see][]{Rees1975,Meszaros1988,Weisskopf2009}. The study of
the state of polarization would unveil the magnetic field and the geometry of the sources and it
would pinpoint the emission processes at work, discriminating among competitive models otherwise
equivalent from the spectral or the timing point of view. This is the case of emission geometry in
pulsars \citep{Dyks2004} or X-ray pulsars in binaries \citep{Meszaros1988}, but peculiar signatures
are also expected for isolated neutron stars because of the different opacity of the two normal
modes in a magnetized plasma and because of vacuum polarization
\citep{Canuto1971,Pavlov2000,Lai2002,Heyl2003}. Moreover polarimetry is a powerful probe to
investigate fundamental theories. General Relativity in the strong field regime can be tested by
means of the rotation of the plane of polarization with energy expected for stellar-mass
black-holes, and the amplitude of the effect would provide a measurement of the spin
\citep{Stark1977,Dovciak2008,Li2009}. Instead a rotation of the polarization angle increasing with
distance could tightly constrain the vacuum birefringence expected in some theories of Quantum
Gravity \citep{Gambini1999,Mitrofanov2003,Kaaret2004}.

As a proof for the impelling interest in X-ray polarimetry, many authors have attempted to
extract the polarization information as a byproduct of existing imaging devices. By selecting
those events which are scattered and detected between two adjacent pixels, any imaging instrument is
in principle a Compton polarimeter because the line connecting the hit pixels approximates the
scattering direction. While some results have been achieved for the Crab Nebula with INTEGRAL
\citep{Dean2008,Forot2008}, measurements of this kind may be affected by strong systematic
effects, are limited to strong sources because of the low Compton scattering probability in the
detector and, in many case, remains questionable \citep{Coburn2003,Rutledge2004}.

Today gas detectors able to image the tracks of photoelectrons provide a valuable alternative to
classical techniques. Basically, they promise a jump in sensitivity thanks to the much higher
capability to collect photons, obtained with a larger energy band with respect to Bragg
polarimeters and a lower energy threshold than Compton instruments. In the following we present the
Gas Pixel Detector (GPD hereafter), one of the first devices able to resolve photoelectron paths
in a gas mixture at atmospheric pressure even at low energy and specifically designed for
astrophysical application.

\section{Photoelectric effect}

Photoelectric effect is a good analyzer of polarization and a perfect one in case of absorption of
spherically symmetric electron orbitals. The differential cross section of the interaction for
the K-shell is \citep{Heitler1954}:
\begin{equation}
\frac{d\,\sigma_{ph}^K}{d\,\Omega}=r_0^2\alpha^4
Z^5\left(\frac{m_e c^2}{E}\right)^{\frac{7}{2}}\frac{4\sqrt{2 } \sin
^2\theta\cos^2\phi}{\left(1-\beta\cos\theta\right)^4}\propto\frac{\sin
^2\theta\cos^2\phi}{\left(1-\beta\cos\theta\right)^4}, \label{eq:PhCrSecDiff}
\end{equation} 
where $\beta$ is the photoelectron velocity in units of $c$ and $r_0$ is the classical electron
radius. The angles $\theta$ and $\phi$ are those that the initial direction of the photoelectron
makes with the direction of the absorbed photon and its electric field, respectively. The
probability of photoelectron emission in a certain azimuthal direction $\phi$ is modulated as a
$\cos^2$ function and hence polarized photons cause a modulation in the direction of ejection, the
peak corresponding to the direction of polarization.

If other electronic shells are considered, the lack of spherical symmetry requires to include in
Equation~(\ref{eq:PhCrSecDiff}) an asymmetry factor $b$ \citep{Ghosh1983}:
\begin{equation}
\frac{d\,\sigma_{ph}}{d\,\Omega}=\frac{\sigma_{ph}^{tot}}{4\pi}\left[1+\frac{b}{2}\left(\frac{
3\sin^2\theta\cos^2\phi } {\left(1+\beta\cos\theta\right)^4}-1\right)\right].
\label{eq:PhCrSecDiffGh}
\end{equation}
The value of $b$ is a function of energy \citep{Yeh1993} but in general $b\leq2$, where the
equality holds for spherical shells and hence Equation~\ref{eq:PhCrSecDiffGh} collapses to
Equation~\ref{eq:PhCrSecDiff}.

In the following we will assume that the $\cos^2$ modulation is complete for completely polarized
photons since photoelectric polarimeters in general work above the K-shell energy threshold and
this makes negligible the contribution of not spherical shells. For example, in the working band of
the GPD 99\% of photons are absorbed by 1s or 2s orbitals, which are completely modulated.

\section{The Gas Pixel Detector}

\begin{figure}[tbp]
\begin{center}
\subfloat[\label{fig:GPD_principle}]{\includegraphics[angle=0,totalheight=4.4cm]{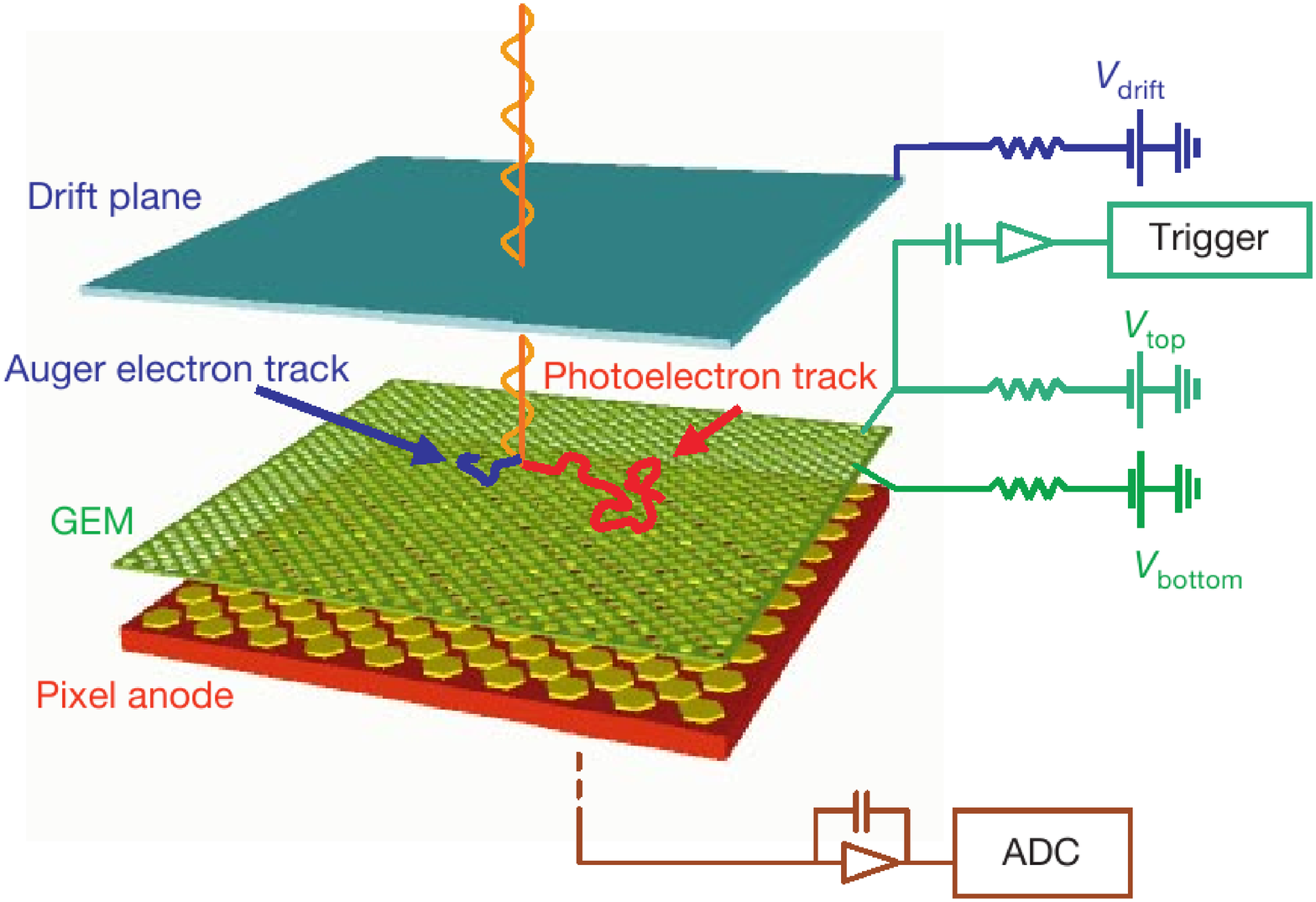}}
\hspace{2mm}
\subfloat[\label{fig:Track52}]{\includegraphics[angle=0,totalheight=4.4cm]{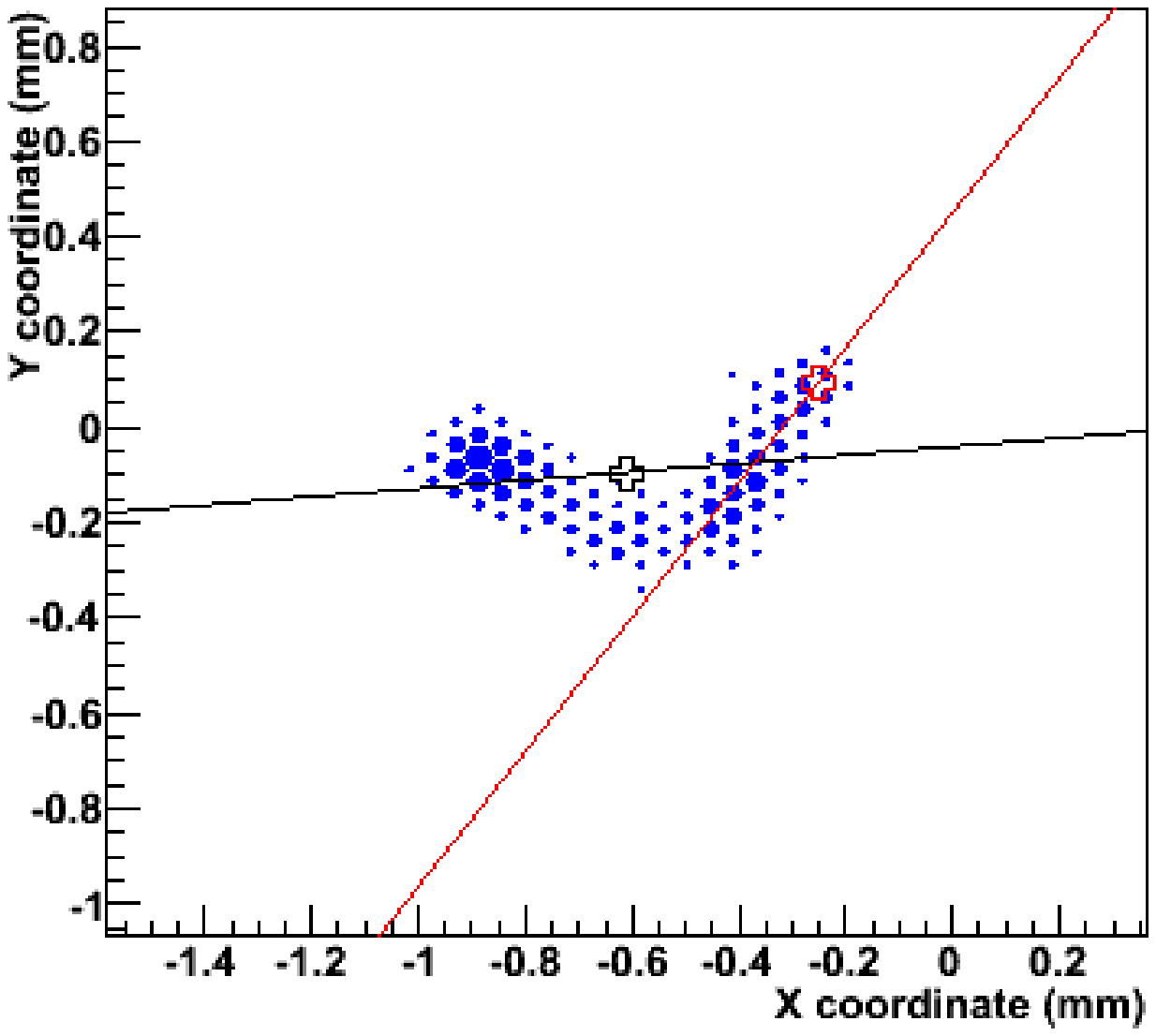}}
\end{center}
\caption{({\bf a}) Principle of operation of the Gas Pixel Detector \citep{Costa2001}. ({\bf b})
An example of a real track at 5.2~keV. The reconstructed direction of emission and the impact point
are the red line and the red cross respectively.}
\end{figure}

The Gas Pixel Detector has been developed and is continuously improved by INFN-Pisa in
collaboration with INAF/IASF-Rome \citep{Costa2001,Bellazzini2006,Bellazzini2007}. It is composed
of a sealed gas cell, 1 or 2~cm thick, enclosed by a 50~$\mu$m beryllium window, a GEM \citep[Gas
Electron Multiplier,][]{Sauli1997} and a finely subdivided pixelized detector (see
Figure~\ref{fig:GPD_principle}). When an X-ray photon is absorbed in the gas cell, a photoelectron
is ejected, propagates in the gas loosing energy by ionization and is eventually stopped after a
path of a few hundreds of microns (track). Alongside the photoelectron it is likely the emission of
an Auger electron: the latter must be distinguished since its direction of ejection is isotropic and
doesn't bring memory of polarization. This implies that a hard limit to the energy threshold is
about twice the binding K-shell energy of the absorbing gas component because in this case the track
of photoelectron is more energetic and hence longer. The electrons produced along the path of the
photoelectron (and of the Auger electron) drift in the gas and are amplified thanks to the electric
field generated by the GEM. Eventually the charge distribution is collected on the pixellated top
metal layer of a 15$\times$15~mm$^2$ VLSI ASIC which provides a true 2D image of the photoelectron
track. This latter component, realized in 0.18~$\mu$m CMOS technology, is the actual breakthrough of
the instrument: the 105,600 pixels are arranged in hexagonal pattern at 50~$\mu$m pitch and are
connected to full and independent electronics chains, which are built immediately below the top
layer of the ASIC and includes pre-amplifier, shaping amplifier, sample and hold and multiplexer. An
important characteristic of the ASIC is that it has self-triggering capabilities which allow to
read-out only a small region enclosing those pixels which were actually hit. The whole
300$\times$352 matrix is divided into 16 (or 8) clusters, each further subdivided into mini-clusters
of 4 pixels. When the charge collected by a mini-cluster is higher than a threshold of $\sim$2000
electrons, corresponding to a few electrons before GEM amplification, a trigger is generated and the
rectangular region of 10 or 20 pixels (externally selectable) around the mini-cluster(s) which
triggered, called Region Of Interest (ROI), is fetched and read-out. Immediately after the event
(delay of a few $\mu$s), the pedestals in the ROI are read-out (one or more times) and subtracted to
data. The electronic noise of the pixels is only 50 electrons ENC \citep{Bellazzini2006}.

The first part of the track produces a lower charge density than the final one, since energy losses
by ionization are inversely proportional to the energy $E$, but the former contains the large part
of the information on polarization. Elastic scatterings, occurring with a probability $\propto
Z^2/E^2$ ($Z$ is the average atomic number of the mixture), isotropize the direction of
photoelectrons which eventually looses any correlation with the polarization. Hence is of
fundamental importance to combine a good quantum efficiency and the capability to resolve and
distinguish the initial part of the track. The small pixel size of the ASIC and the low diffusion of
the mixtures we used, composed of helium, neon or argon and dimethyl ether (DME hereafter), assure
that even at a few keV the photoelectron tracks are well resolved. Moreover we have developed an
algorithm which can effectively reconstruct the initial direction of emission and the impact point
by analyzing only the initial part of the photoelectron path \citep{Bellazzini2003}. This is
distinguished by those hit pixels which are at the edge of the track with the lower ionization
density: by a proper selection and weight of the contribution of each pixel, the best estimate of
the direction of emission and of the absorption point are derived as the direction which maximizes
the second moment and the center of mass of the charge distribution, respectively.

An image of a real track at 5.2~keV in pure DME at 0.8~bar is reported in Figure~\ref{fig:Track52}.
The effect of a large scattering is clearly resolved as the denser region where photoelectron is
eventually stopped (Bragg peak). The discrimination of the first part of the track allows for a
reconstruction of the direction of emission and of the  absorption point (depicted as the line and
the cross in red) by far better than if the entire track is analyzed (in black in Figure). The
possibility to measure the impact point, with a resolution of the order of 150~$\mu$m, gives to the
GPD the unique possibility among gas polarimeters to image the source and reduce the background in
the focus of a telescope to a negligible level. Moreover, as any conventional gas counter, the
energy and the time of the event are also available. Goal spectroscopic and timing resolutions are
20\% at 5.9~keV and $\sim$8~$\mu$s, achieved by reading-out the signal of the GEM together with
that of the ASIC. However even current spectral performance, obtained by summing the charge collected in
hit pixels, is already encouraging, 24\% at 5.9~keV \citep{Muleri2009b}.

\begin{figure}[tbp]
\begin{center}
\includegraphics[angle=0,totalheight=4.2cm]{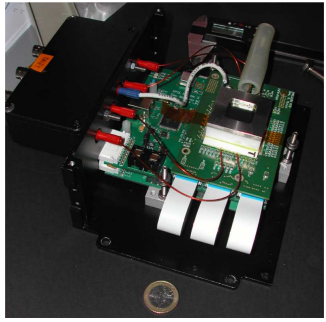}
\end{center}
\caption{A picture of the GPD detector and of the back-end electronics.}
\label{fig:GPD}
\end{figure}

The GPD is assembled with low-outgassing materials in collaboration with Oxford Instruments
Analytical Oy and it doesn't show any significant degradation after more than 1~year of continuous
operation. The gas cell is sealed and this allows to build a very compact instrument: the detector
and the back-end electronics currently in use in our laboratory are in a box
140$\times$190$\times$70~mm$^3$ which weights 1.6~kg and the power consumption is $\leq$5~W (see
Figure~\ref{fig:GPD}). Notwithstanding the mixture can be refilled to test different configurations
in view of the great importance of this choice. Currently, best performance in the energy range
2-10~keV, where first use is expected, is achieved with a mixture of helium and DME at 1~atm, but we
are also studying mixtures of argon to extend the energy range up to $\sim$35~keV (see
Section~\ref{sec:SatMis}).

\begin{figure}[tbp]
\begin{center}
\subfloat[\label{fig:ModPol}]{\includegraphics[angle=0,totalheight=4.1cm]{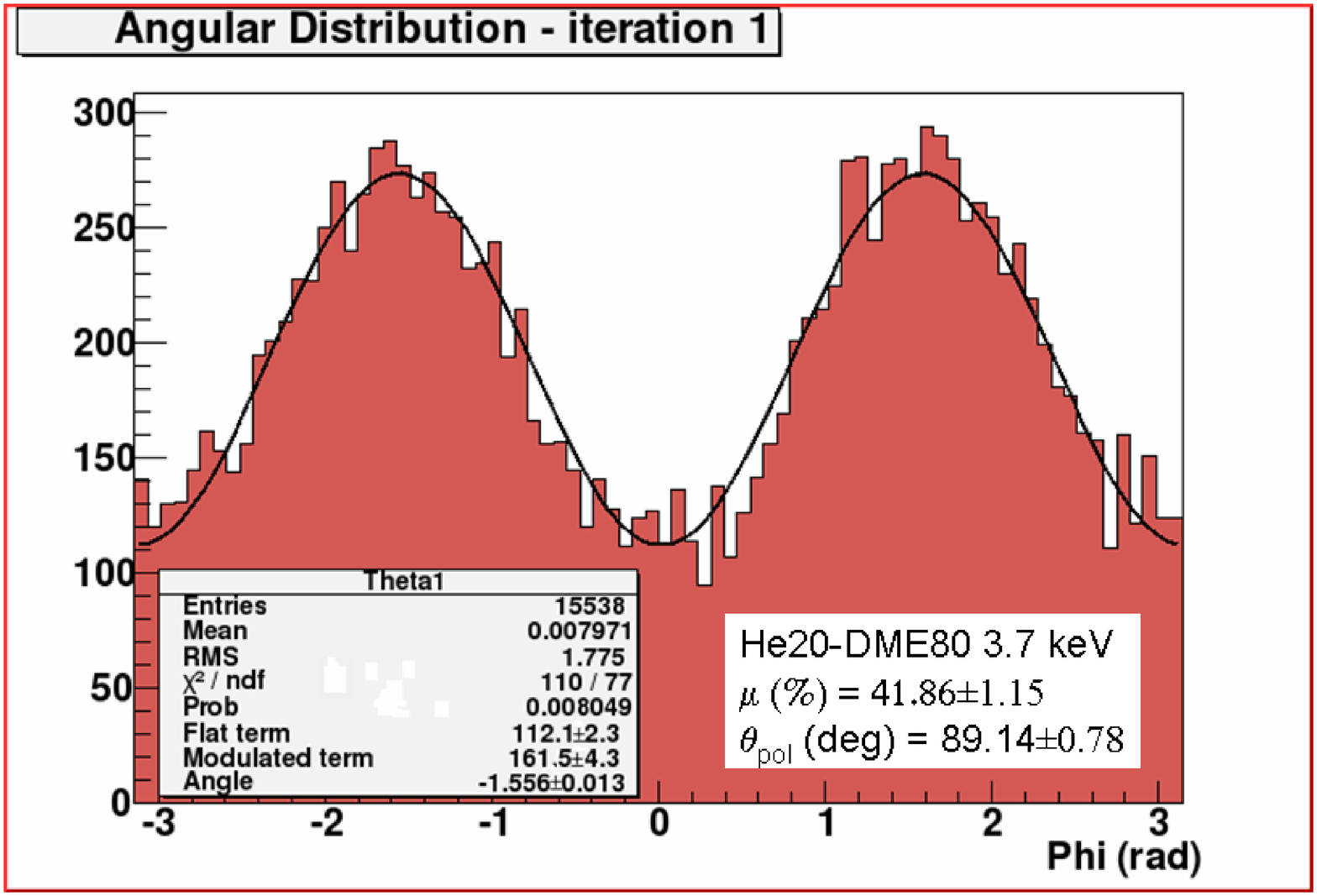}}
\hspace{2mm}
\subfloat[\label{fig:MuVsEn}]{\includegraphics[angle=0,totalheight=4.1cm]{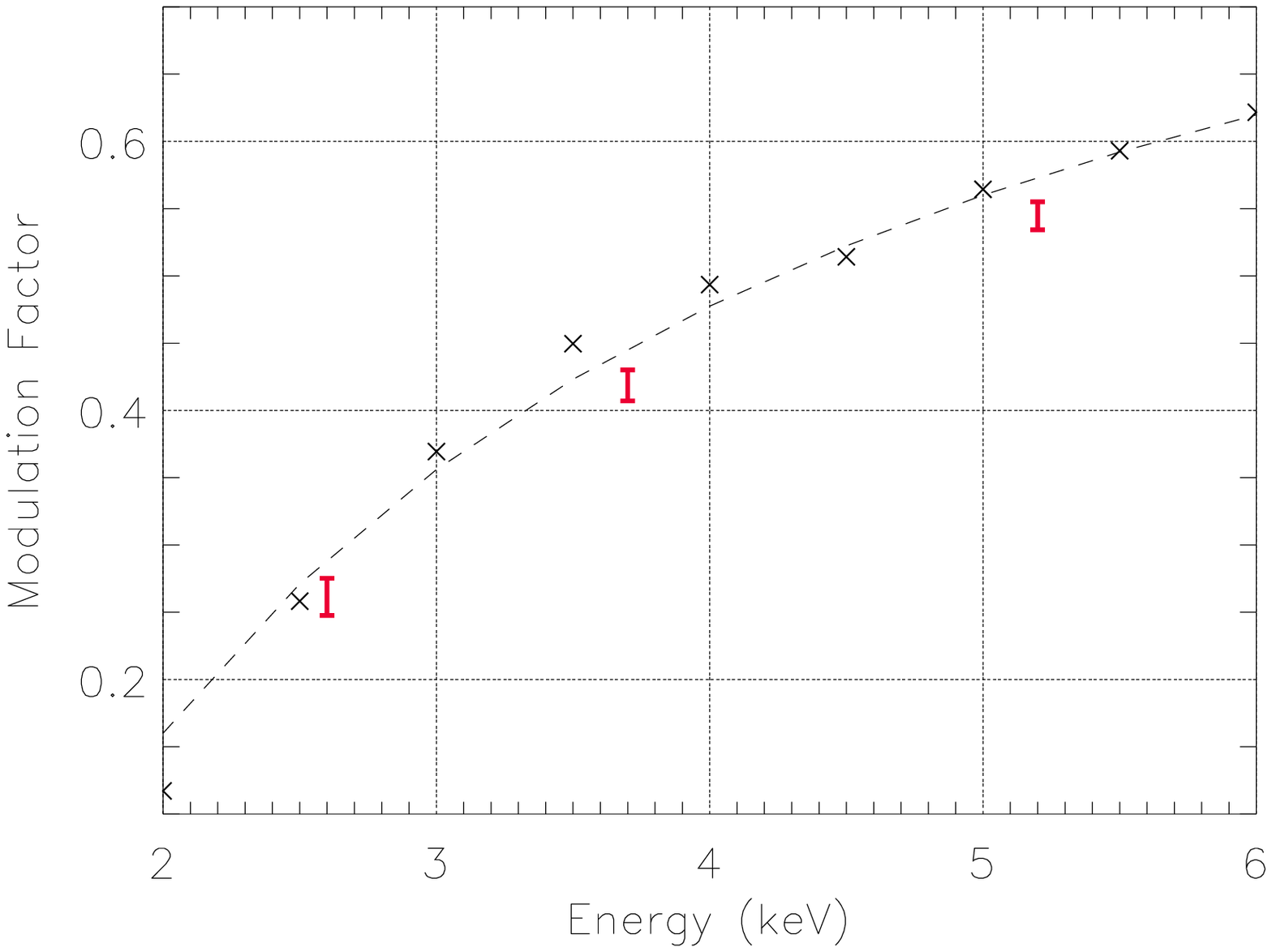}}
\end{center}
\caption{({\bf a}) Modulation curve for a mixture of helium and DME and completely polarized
photons at 3.7~keV. ({\bf b}) Modulation factor measured at 2.6, 3.7 and 5.2~keV (red points)
compared to the values expected from Monte Carlo simulations (crosses). The dashed line is a fit to
Monte Carlo data \citep[both Figures after][]{Muleri2008}.}
\end{figure}

An example of the histogram of emission directions as a function of the azimuthal angle (modulation
curve) is shown in Figure~\ref{fig:ModPol} for 3.7~keV polarized photons. The amplitude of the
modulation measured at 2.6, 3.7 and 5.2~keV for completely polarized photons (modulation factor) is
consistent with the values expected on the basis of Monte Carlo simulations (see
Figure~\ref{fig:MuVsEn}).

\section{Satellite missions} \label{sec:SatMis}

There aren't critical issues for the use of GPD on-board space satellites and the instrument has
a very high readiness level. The VLSI technology naturally withstands high levels of radiation
and tests at the Heavy Ion Medical Accelerator in Chiba (HIMAC) confirmed that performance remains
nominal after the irradiation of iron ions corresponding to several years in orbit. There is a long
heritage in avoiding and eventually controlling pollution of sealed gases on-board space
satellites. The detector survived vibrations and thermo-vacuum tests between -15 and 45$^\circ$C and
both the GEM and the 50~$\mu m$ thin beryllium window are compatible with the use in space. The gas
amplification gain is maintained low thanks to the low pixels noise and this reduces the
probability of destructive discharges.

The imaging capability of the GPD will be exploited at best with the use of an X-ray optics. This
will assure (i) an adequate collecting area, (ii) the possibility to resolve extended sources
and (iii) to reduce to a negligible level the background, expected to be more than two orders of
magnitude lower than the fainter source accessible to X-ray polarimetry. The intrinsic azimuthal
symmetry of the instrument makes the systematic effects much lower than 1\%, Bellazzini et al.
\citep{Bellazzini2010} reported that for $^{55}$Fe 5.9~keV unpolarized photons the residual
modulation is 0.18$\pm$0.14\%. This allows to avoid the rotation of the instrument which is an
odd complication to the design of modern three-axis stabilized satellites.

The GPD is currently included in several proposals of mission which have passed a certain level of
selection (see Table~\ref{tab:Missions}). One possibility is \emph{XPOL} on-board the large
satellite \emph{International X-ray Observatory} (IXO), which allows to exploit a huge collecting
area (1.5~m$^2$ at 3~keV) and achieve a fine angular resolution, 5~arcsec including the blurring due
to inclined penetration of photons in the gas cell. Even if only a small fraction of the observation
time would be dedicated to XPOL, a minimum detectable polarization (MDP) of 1\%  at the 99\%
confidence level could be reached within 100~ks for 1~mCrab source, thus making possible polarimetry
even of faint extragalactic sources \citep{Costa2008,Bellazzini2010b}. Nevertheless the sensitivity
of the GPD would allow to achieve a wealth of results even with a less demanding mission profile.
Using a telescope of modest area dedicated to polarimetry, tens of galactic and extragalactic bright
sources would be accessible to the level of 1\% in a few days of observation. Such a small missions
could be placed in orbit with a moderate cost and well before IXO, whose launch is expected in 2021,
with the additional advantage to drive the development and the observations of the latter mission.

\begin{table}
\centering
\begin{tabular}{rccccc}
& \multicolumn{2}{c}{\bf NHXM} & {\bf POLARIX} & {\bf HXMT} & {\bf XPOL} \\
\hline
Energy range (keV) & 2-10 & 6-35 & 2-10 & 2-10 & 2-10 \\
$\delta$E at 6~keV & 20\% & $<$20\% & 20\% & 20\% & 20\% \\
MDP in 100~ks (mCrab) & 3\% (10) & 4\% (10) & 3.6\% (10) & 3\% (10) & 1\% (1) \\
Ang. Res. (arcsec) & 15 & 22 & 25 & 40 & 5 \\
fov (arcmin) & \multicolumn{2}{c}{12$\times$12} & 15$\times$15 & 25$\times$25 & 2.6$\times$2.6 \\
Timing resolution ($\mu$s) & \multicolumn{2}{c}{8} & 8 & 8 & 8 \\
Telescopes & \multicolumn{2}{c}{1} & 3 & 2 & 1 \\
Focal length (m) & \multicolumn{2}{c}{10} & 3.5 & 2.1 & 20 \\
Area at 3~keV (cm$^2$) & \multicolumn{2}{c}{520} & 380 & 610 & 14300 \\
\hline
\end{tabular}
\caption{Comparison of missions with the GPD on-board. The MDP is referred to a 100~ks observation
of a source with a flux corresponding to the value reported in parentheses. In the 2-10~keV energy
band, 1~mCrab corresponds to a flux $2.3\;10^{-11}$~erg/(s cm$^2$), while it is
$2.5\;10^{-11}$~erg/(s cm$^2$) between 6 and 35~keV.}
\label{tab:Missions}
\end{table}

To the pathfinder mission profile belong missions like \emph{POLARIX} \citep{Costa2006,Costa2009},
an Italian mission dedicated to X-ray polarimetry between 2 and 10~keV which concluded a phase~A
study at the end of 2008 and it's currently waiting for the possible selection to launch by the
Italian Space Agency (ASI), and the polarimeters on-board the Chinese mission \emph{HXMT}
\citep{Soffitta2008}. These missions (see Table~\ref{tab:Missions}) exploit a small cluster of
identical telescopes to reduce costs. While summing data from more detectors doesn't impact on
sensitivity thanks to the negligible background, the replica of identical optics allows to achieve
an equivalent area with a smaller number of shells, whose production compose the large part of the
telescope costs.

\begin{figure}[tbp]
\begin{center}
\includegraphics[angle=0,totalheight=4.1cm]{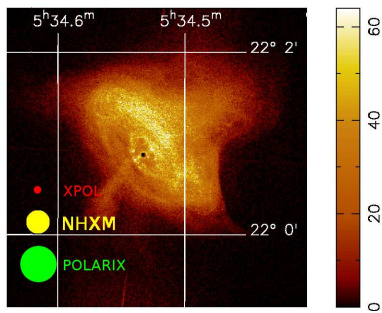}
\end{center}
\caption{Angular resolution of POLARIX, NHXM and IXO compared with the Crab Nebula \citep[background
image from][]{Weisskopf2000}.}
\label{fig:CrabNebula}
\end{figure}

A further interesting possibility which is emerging in Italy is the \emph{New Hard X-ray
Mission} (NHXM hereafter) designed to extend up to 80~keV the use of fine angular resolution
multilayer optics. The peculiarity of this mission is to perform contemporaneously imaging and
spectroscopy with three identical telescopes and stacked detectors and polarimetry with a fourth
optics. Two GPDs filled with mixtures of Helium/DME and Argon/DME cover the energy band 2-10~keV and
6-35~keV with a similar sensitivity, 3\% and 4\% for a 10~mCrab source in 100~ks respectively, and
are placed alternatively in the focus of the telescope dedicated to polarimetry. The field of view
is 6$\times$6~arcmin and the angular resolution is 15~arcsec below 10~keV and 22~arcsec at 30~keV,
sufficient to resolve main structures of extended sources like the Crab Nebula (see
Figure~\ref{fig:CrabNebula}).

\section{Conclusions}

X-ray polarimetry is a fundamental tool for increasing our knowledge of emission processes acting
in compact sources and for investigating theories of fundamental physics. Today gas detectors
which resolve photoelectron tracks provide a valuable alternative to classical techniques of Bragg
diffraction and Thomson scattering and promise to perform polarimetry of tens of galactic sources
even within a small mission to be launched in a few years. The Gas Pixel Detector is in an advanced
phase of development: the sensitivity expected on the basis of Monte Carlo simulations is confirmed
by measurements and the detector survived irradiation, vibrations and thermo-vacuum tests. Once
in-orbit with an X-ray telescope, the GPD will contemporaneously measure the polarization, the
energy, the time and the direction of arrival of photons. Measurements at the level of 1\% and below
are within the possibility of the GPD because systematic effects are well under control thanks
to the intrinsic azimuthal symmetry of the instrument. On-board of small missions, the GPD could
reach between 2 and 10~keV a minimum detectable polarization of about 3\% for 10~mCrab sources in
100~ks of observation and NHXM offers the possibility to extend polarimetry up to 35~keV with a
similar sensitivity. The GPD will also be part of the scientific payload of the large mission IXO,
whose large collecting area will allow for the measurements of polarization even of faint
extragalactic sources (1\% for 1~mCrab source in 100~ks) with a fine angular resolution.

\section*{Acknowledgments} 

The activity was supported by ASI contracts I/088/060 and I/012/08/0. The authors acknowledge the
GPD team for useful discussions and suggestions.

\bibliographystyle{elsarticle-harv}
\bibliography{ReferencesShort}

\begin{thebibliography}{39}
\expandafter\ifx\csname natexlab\endcsname\relax\def\natexlab#1{#1}\fi
\expandafter\ifx\csname url\endcsname\relax
  \def\url#1{\texttt{#1}}\fi
\expandafter\ifx\csname urlprefix\endcsname\relax\def\urlprefix{URL }\fi

\bibitem[{{Bellazzini} and {Spandre}(2010)}]{Bellazzini2010}
{Bellazzini}, R., {Spandre}, G., 2010. {Photoelectric polarimeters}. In: X-ray
  Polarimetry: A New Window in Astrophysics. Cambridge University Press.

\bibitem[{{Bellazzini et al.}(2003)}]{Bellazzini2003}
{Bellazzini et al.}, 2003. In: Proc. of SPIE. Vol. 4843. p. 383.

\bibitem[{{Bellazzini et al.}(2006)}]{Bellazzini2006}
{Bellazzini et al.}, 2006. NIMA 566, 552.

\bibitem[{{Bellazzini et al.}(2007)}]{Bellazzini2007}
{Bellazzini et al.}, 2007. NIMA 579, 853.

\bibitem[{{Bellazzini et al.}(2010 in press)}]{Bellazzini2010b}
{Bellazzini et al.}, 2010 in press. {A polarimeter for IXO}. In: X-ray
  Polarimetry: A New Window in Astrophysics.

\bibitem[{{Canuto et al.}(1971)}]{Canuto1971}
{Canuto et al.}, 1971. \prd 3, 2303.

\bibitem[{{Coburn} and {Boggs}(2003)}]{Coburn2003}
{Coburn}, W., {Boggs}, S.~E., 2003. \nat 423, 415.

\bibitem[{{Costa et al.}(2001)}]{Costa2001}
{Costa et al.}, 2001. \nat 411, 662.

\bibitem[{{Costa et al.}(2006)}]{Costa2006}
{Costa et al.}, 2006. In: Proc. of SPIE. Vol. 6266. p. 62660R.

\bibitem[{{Costa et al.}(2008)}]{Costa2008}
{Costa et al.}, 2008. In: Proc. of SPIE. Vol. 7011. pp. 70110F--1.

\bibitem[{{Costa et al.}(2009)}]{Costa2009}
{Costa et al.}, 2009. {POLARIX: a pathfinder mission of X-ray polarimetry}. In
  preparation.

\bibitem[{{Dean et al.}(2008)}]{Dean2008}
{Dean et al.}, 2008. Science 321, 1183.

\bibitem[{{Dov{\v c}iak et al.}(2008)}]{Dovciak2008}
{Dov{\v c}iak et al.}, 2008. \mnras 391, 32.

\bibitem[{{Dyks et al.}(2004)}]{Dyks2004}
{Dyks et al.}, 2004. \apj 606, 1125.

\bibitem[{{Forot et al.}(2008)}]{Forot2008}
{Forot et al.}, 2008. \apjl 688, L29.

\bibitem[{{Gambini} and {Pullin}(1999)}]{Gambini1999}
{Gambini}, R., {Pullin}, J., 1999. \prd 59~(12), 124021.

\bibitem[{{Ghosh}(1983)}]{Ghosh1983}
{Ghosh}, P.~K., 1983. {Introduction to Photoelectron Spectroscopy}. John Wiley
  \& Sons.

\bibitem[{{Heitler}(1954)}]{Heitler1954}
{Heitler}, W., 1954. {Quantum theory of radiation}. International Series of
  Monographs on Physics, Oxford: Clarendon, 1954, 3rd ed.

\bibitem[{{Heyl et al.}(2003)}]{Heyl2003}
{Heyl et al.}, 2003. \mnras 342, 134.

\bibitem[{{Hughes et al.}(1984)}]{Hughes1984}
{Hughes et al.}, 1984. \apj 280, 255.

\bibitem[{{Kaaret}(2004)}]{Kaaret2004}
{Kaaret}, P., 2004. \nat 427, 287.

\bibitem[{{Lai} and {Ho}(2002)}]{Lai2002}
{Lai}, D., {Ho}, W.~C.~G., 2002. \apj 566, 373.

\bibitem[{{Li et al.}(2009)}]{Li2009}
{Li et al.}, 2009. \apj 691, 847.

\bibitem[{{Long et al.}(1979)}]{Long1979}
{Long et al.}, 1979. \apjl 232, L107.

\bibitem[{{Meszaros et al.}(1988)}]{Meszaros1988}
{Meszaros et al.}, 1988. \apj 324, 1056.

\bibitem[{{Mitrofanov}(2003)}]{Mitrofanov2003}
{Mitrofanov}, I.~G., 2003. \nat 426, 139.

\bibitem[{{Muleri et al.}(2008)}]{Muleri2008}
{Muleri et al.}, 2008. NIMA 584, 149.

\bibitem[{{Muleri et al.}(2009)}]{Muleri2009b}
{Muleri et al.}, 2009. JINST 11, 2.

\bibitem[{{Novick et al.}(1972)}]{Novick1972}
{Novick et al.}, 1972. \apjl 174, L1.

\bibitem[{{Pavlov} and {Zavlin}(2000)}]{Pavlov2000}
{Pavlov}, G.~G., {Zavlin}, V.~E., 2000. \apj 529, 1011.

\bibitem[{{Rees}(1975)}]{Rees1975}
{Rees}, M.~J., 1975. \mnras 171, 457.

\bibitem[{{Rutledge} and {Fox}(2004)}]{Rutledge2004}
{Rutledge}, R.~E., {Fox}, D.~B., 2004. \mnras 350, 1288.

\bibitem[{{Sauli}(1997)}]{Sauli1997}
{Sauli}, F., 1997. NIMA 386, 531.

\bibitem[{{Soffitta et al.}(2008)}]{Soffitta2008}
{Soffitta et al.}, 2008. In: Proc. of SPIE. Vol. 7011. pp. 701128--1.

\bibitem[{{Stark} and {Connors}(1977)}]{Stark1977}
{Stark}, R.~F., {Connors}, P.~A., 1977. \nat 266, 429.

\bibitem[{{Weisskopf et al.}(1978)}]{Weisskopf1978}
{Weisskopf et al.}, 1978. \apjl 220, L117.

\bibitem[{{Weisskopf et al.}(2000)}]{Weisskopf2000}
{Weisskopf et al.}, 2000. \apjl 536, L81.

\bibitem[{{Weisskopf et al.}(2009)}]{Weisskopf2009}
{Weisskopf et al.}, 2009. In: Neutron Stars and Pulsars, Astrophysics and Space
  Science Library, Springer Berlin Heidelberg. Vol. 357. p. 589.

\bibitem[{{Yeh}(1993)}]{Yeh1993}
{Yeh}, J.~J., 1993. {Atomic Calculation of Photoionization Cross--Sections and
  Asymmetry Parameters}. Gordon and Breach Science Publishers, Langhorne, PE
  (USA).

\end{thebibliography}

\end{document}